\documentclass[aps,reprint,unsortedaddress,showpacs,floatfix,nofootinbib]{revtex4-1}
\usepackage[colorlinks=true,breaklinks=false]{hyperref}

\usepackage{graphicx}
\usepackage[normalem]{ulem}
\usepackage{amsmath,amstext,amssymb}
\usepackage{dcolumn}
\usepackage{bm}
\usepackage[english]{babel}
\usepackage{xcolor}

\newcommand{\FT}{{\mathcal F}}
\newcommand{\IFT}{{\mathcal F^{-1}}}
\newcommand{\mic}{~{\rm \mu m}}
\newcommand{\imm}{~{\rm mm^{-1}}}
\begin{document}

\title{Soliton-induced nonlocal resonances observed through high-intensity tunable spectrally compressed second-harmonic peaks}

\author{Binbin Zhou,  Hairun Guo,  Morten Bache$^*$}

\affiliation{
DTU Fotonik, 
Technical University of Denmark,
  Bld. 343, DK-2800 Kgs. Lyngby, Denmark.\\
$^*$Corresponding author: moba@fotonik.dtu.dk, \today }

\begin{abstract}
\noindent
Experimental data of femtosecond thick-crystal second-harmonic generation shows that when tuning away from phase matching, a dominating narrow spectral peak appears in the second harmonic that can be tuned over 100's of nm by changing the phase-mismatch parameter. Traditional theory explains this as phase matching between a sideband in the broadband pump to its second-harmonic. However, our experiment is conducted under high input intensities and instead shows excellent quantitative agreement with a nonlocal theory describing cascaded quadratic nonlinearities. This theory explains the detuned peak as a nonlocal resonance that arises due to phase-matching between the pump and a detuned second-harmonic frequency, but where in contrast to the traditional theory the pump is assumed dispersion-free. As a soliton is inherently dispersion-free, the agreement between our experiment and the nonlocal theory indirectly proves that we have observed a soliton-induced nonlocal resonance. The soliton exists in the self-defocusing regime of the cascaded nonlinear interaction and in the normal dispersion regime of the crystal, and needs high input intensities to become excited. 
\end{abstract}

\pacs{42.65.Re, 42.65.Tg, 42.65.Ky, 42.65.Sf}

\maketitle

\begin{figure}[htb]
\centerline{\includegraphics[width=8.4cm]{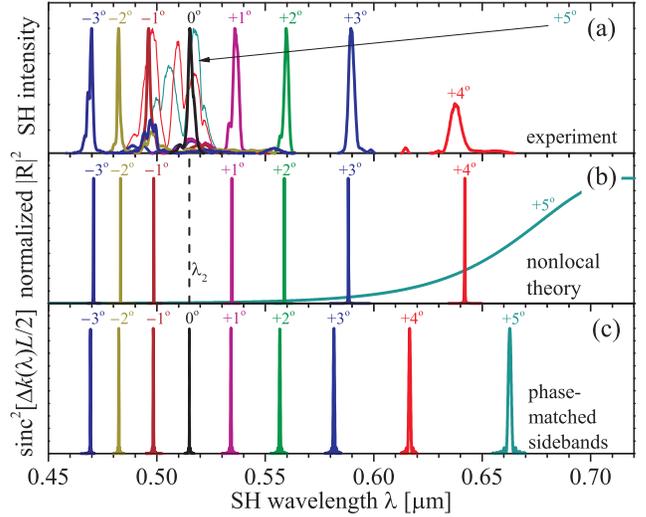}}
\caption{(Color online) (a) Experimental SH normalized spectra sweeping the phase-mismatch parameter from negative to positive values (external angle tuned from $-3^\circ$ to $+5^\circ$ around $\Delta k=0$). An $L=25$ mm BBO crystal was used pumped with $\lambda_1=1.03\mic$ 58 fs FWHM 80 GW/cm$^2$ pulses. (b) The nonlocal theory, predicting $I_2\propto|R(\Omega)|^2$. (c) Phase-matched sidebands theory, predicting $I_2\propto {\rm sinc^2}[\Delta k(\lambda) L/2]$. 
}
\label{fig:SH-Dk}
\end{figure}

\noindent
A common observation in second-harmonic generation (SHG) of broadband laser pulses in thick crystals, is that when a phase mismatch $\Delta k$ is imposed, the second harmonic (SH) spectrum is dominated by a spectrally compressed peak that is wavelength-tunable through $\Delta k$ \cite{bakker:1992,Rassoul:1997,Pioger:2002,Zhu:2004, baronio:2004,Moutzouris2006,Marangoni:2007,Pontecorvo2011}. Figure \ref{fig:SH-Dk}(a) shows data from an experiment we performed using a thick $\beta$-barium borate (BBO) crystal. The input fundamental wave (FW, center frequency $\omega_1$) was an intense femtosecond pulse loosely focused and collimated at the crystal entrance to avoid diffraction. The tuning around $\Delta k=0$ was achieved by rotating the external angle of the crystal. A striking wavelength tunability over 100's of nanometers is possible, and the peak is also strongly compressed compared to the ideal thin-crystal bandwidth (in this case about 40 nm FWHM). The compressed SH peak pertains for large negative tuning angles, while it disappears for large positive tuning angles (here $+5^\circ$). The spectral compression is traditionally explained by a \textit{phase-matched sidebands theory}, which uses the classical result that the SH efficiency $\propto {\rm sinc}^2[\Delta kL/2]$: this explains the decreasing bandwidth in a thick crystal, and the frequency dependence of $\Delta k(\omega)=k_2(\omega)-2k_1(\omega/2)$ explains how a SH sideband frequency strongly detuned from the degenerate SH frequency $\omega_2=2\omega_1$ can become phase matched when $\Delta k(\omega_2)\neq 0$. Figure \ref{fig:SH-Dk}(c) shows the predicted phase-matching wavelengths by the phase-matched sidebands theory, and remarkably it cannot explain the experimental data for large positive tuning angles.

Instead an alternative \textit{nonlocal theory}, shown in Fig. \ref{fig:SH-Dk}(b), predicts "resonance" wavelengths in the nonlocal response function that up to $+4^\circ$ are in excellent agreement with the experimental peak positions. At $+5^\circ$ the phase-matching condition behind this resonance is no longer fulfilled, yielding a broadband, non-resonant nonlocal response that does not favor the formation of a detuned SH peak (see also the numerical results presented below). This agrees with the experimental data that shows no detuned SH peak there, and instead the radiation at zero detuning (around $\lambda_2$) dominates. The nonlocal theory predicts that under strongly phase-mismatched (cascaded) SHG, the well-known $\chi^{(2)}:\chi^{(2)}$ Kerr-like nonlinearity acts like a temporally "nonlocal" nonlinearity \cite{bache:2007a}. In previous work, this nonlocal theory could predict the detuned SH peak observed under similar conditions as Fig. \ref{fig:SH-Dk}(a) \cite{bache:2008,bache:2010,Guo:2013}.

The purpose of this paper is to understand why our experiment follows the nonlocal theory and investigate the physics behind the two models to elucidate their applicabilities. In brief, the phase-matched sidebands theory takes into account the FW dispersion, while the nonlocal theory assumes a dispersion-free FW. The experimental conditions are carefully chosen so the two cases can be distinguished, and Figs. \ref{fig:SH-Dk} (b) and (c) show that for tuning angles of $+3^\circ$ and beyond they start to differ. Such strongly detuned SH peaks are only observable with high input intensities, which may excite self-defocusing temporal solitons at the FW wavelength: The cascaded SHG leads to a Kerr-like nonlinearity $n_{2, \rm casc}^I\propto -d_{\rm eff}^2/\Delta k$ \cite{Ostrovskii:1967,desalvo:1992}, which is self-defocusing for positive phase mismatch $\Delta k=k_2-2k_1$ (positive tuning angles). Since the FW has normal group-velocity dispersion (GVD) in BBO at $1.03\mic$, a FW self-defocusing soliton can be excited \cite{liu:2000-solitons,ashihara:2002,ashihara:2004,moses:2006,Zeng:2008,moses:2007,zhou:2012} when the self-focusing material Kerr nonlinearity is outbalanced (so $n_{2, \rm casc}^I+n_{2,\rm Kerr}^I<0$). The phase-matching condition between such a soliton and the detuned SH peaks will be accurately described by the nonlocal theory as the soliton is inherently dispersion-free. Thus, the excellent agreement between the nonlocal resonance wavelengths and the experimental data indirectly proves that this is a soliton-induced nonlocal resonance. Conversely, the historical measurements \cite{bakker:1992,Rassoul:1997,Pioger:2002,Zhu:2004, baronio:2004,Moutzouris2006,Marangoni:2007,Pontecorvo2011} used too low intensities for soliton formation: in this case the FW is dispersive and the phase-matched sideband theory prevails.


Cascaded quadratic nonlinearities have been extensively investigated experimentally recently for ultrafast pulse compression and soliton formation \cite{Rassoul:1997,liu:1999,liu:2000-solitons,ditrapani:2001,ashihara:2002,ditrapani:2003,Jedrkiewicz:2003-xwave,ashihara:2004,moses:2006,Zeng:2008,moses:2007,zhou:2012},
supercontinuum generation \cite{Fuji:2005,Langrock:2007,Phillips:2011-ol,zhou:2012}, white light continuum from filaments \cite{Srinivas:2005,Kumar:2007,Kumar:2008}, frequency comb generation \cite{Ulvila:2013}, femtosecond modelocking \cite{Zavelani-Rossi:1998,Qian:1999,Agnesi:2005,Agnesi:2008,Zondy:2010,Meiser:2013,Phillips:2014}, compensation of self-focusing effects \cite{Beckwitt:2001,Conti:2002}, material properties \cite{Cussat-Blanc:1998,Ashihara:2003,Bache:2013,moses:2007a} and ultrafast pulse control \cite{Noordam:1990,Liu:2000,Ashihara:2003a,xu:2004,ilday:2004,Baronio:2006,moses:2006b,moses:2007a,Su:2006,Moutzouris2006,Marangoni:2006,Marangoni:2007,centini:2008,Fazio:2009,Pontecorvo2011,Valiulis:2011,Baronio:2012}.
One particular feature of cascaded quadratic nonlinearities is that they induce a nonlocal nonlinearity, either spatially \cite{nikolov:2003} or temporally \cite{bache:2007a}. In the spatial nonlocal case, the nonlinearity depends not only on the local intensity but also on its neighboring points \cite{snyder:1997} (see \cite{krolikowski:2004} for a review); spatial nonlocal nonlinearities have been observed in various physical environments such as heat conduction \cite{Dabby:1968}, ballistic atomic transport \cite{Skupin:2007}, diffusion \cite{suter:1993}, charge separation \cite{Ultanir:2003-ol}, or long-range particle interaction as in e.g. dipolar Bose-Einstein condensates \cite{burger:1999,*Dalfovo:1999,*Lahaye:2009}, and nematic liquid crystals \cite{braun:1993,*assanto:2003}. We here consider a temporal nonlocal nonlinearity, where the nonlinear cascading response at time $\tau$ (in the moving reference frame) relies not just on the instantaneous (local) field values but also at times before and after that. In time-domain a first-order expansion of the nonlocal response \cite{Guo:2013} reveals the analogy to the cascading-induced controllable pulse self steepening \cite{moses:2006b,moses:2007a}, while a frequency-domain description reveals that the nonlocal response can be either non-resonant (i.e. ultrabroadband, which is the optimal situation for few-cycle pulse compression \cite{bache:2007a,bache:2008,zhou:2012}) or resonant as in the case investigated here. The first observation of a temporal cascaded self-defocusing soliton \cite{ashihara:2002} was actually carried out in the resonant regime (see \cite{bache:2007a}), but this was prior to the discovery of a nonlocal resonance regime for such an interaction \cite{bache:2007a} and the SH spectrum was not recorded to document the connection between soliton formation and the SH spectral resonance. The spatial equivalent of a resonant nonlocal nonlinearity occurs when the SH experiences negative diffraction or when the phase mismatch is negative \cite{nikolov:2003}, but the analytical soliton solutions are very elusive \cite{buryak:2002,esbensen:2012a} due to an oscillatory nature of the nonlocal response in time/space domain that comes as a consequence of the spectral resonance. To our knowledge, a soliton-induced nonlocal resonance, spatially or temporally, has yet to be experimentally observed.


The theory starts with the plane-wave SHG equations
\begin{align}\label{eq:shg-fw-a1}
   \left[i\tfrac{\partial}{\partial z}- \tfrac{1}{2}k_1^{(2)}\tfrac{\partial^2}{\partial \tau^2}\right]E_1=-\tfrac{\omega_1
   d_{\rm eff}}{cn_1} E_1^*E_2 e^{i\Delta k z}\\
\label{eq:shg-sh-a1}
\left[i\tfrac{\partial}{\partial z}-id_{12}\tfrac{\partial}{\partial
    \tau}- \tfrac{1}{2}k_2^{(2)}\tfrac{\partial^2}{\partial \tau^2}\right]E_2=-\tfrac{\omega_1 d_{\rm eff}}{cn_2} E_1^2
e^{-i\Delta k z}
\end{align}
where $k_j(\omega)=n_j(\omega)\omega/c$ are the wave numbers, $n_j(\omega)$ the frequency dependent refractive indices of the FW ($j=1$) and SH ($j=2$), $n_j\equiv n_j(\omega_j)$, $\Delta k=k_2^{(0)}-2k_1^{(0)}$ the phase mismatch parameter, $d_{12}=k_1^{(1)}-k_2^{(1)}$ the group-velocity mismatch (GVM) parameter,  $k_j^{(2)}$ the GVD coefficients. Higher-order dispersion $k_{j}^{(m)}=d^mk_j/d\omega^m|_{\omega=\omega_j}$ is neglected as to allow for analytical solutions, but all the plots use exact (material) dispersion from the Sellmeier equations (taken from \cite{Zhang:2000}). Finally, $d_{\rm eff}$ is the effective $\chi^{(2)}$ nonlinearity. Following \cite{Valiulis:2011}, we take $E_2(z,\tau)=B_2(z,\tau)e^{-i\Delta kz}$, and Eq. (\ref{eq:shg-sh-a1}) in Fourier domain is
\begin{align}\label{eq:SH-FD}
\tfrac{\partial}{\partial z}B_2(z,\Omega)-i D_2(\Omega)B_2(z,\Omega) =i\tfrac{\omega_1 d_{\rm eff}}{cn_2} \FT[E_1^2]
\end{align}
where $D_2(\Omega)= \tfrac{1}{2}k_2^{(2)}\Omega^2-d_{12}\Omega+\Delta k$ is the effective SH dispersion operator. To solve this, we find solutions to the homogeneous equation as $B_2^{(h)}\propto e^{iD_2(\Omega)z}$. A particular solution $B_2^{(p)}$ that is constant in $z$ can be found by requiring that $\FT[E_1^2]$ does not depend on $z$. This yields 
\begin{align}\label{eq:SH-particular}
B_2^{(p)}(\Omega)=-\tfrac{\omega_1 d_{\rm eff}}{cn_2 D_2(\Omega)} \FT[E_1^2]
\end{align}
As the total solution is a linear combination $B_2^{(t)}=B_2^{(h)}+B_2^{(p)}$, appropriate boundary conditions give the homogeneous solution 
\begin{align}\label{eq:SH-homog}
B_2^{(h)}(z,\Omega)=\tfrac{\omega_1 d_{\rm eff}}{cn_2 D_2(\Omega)} \FT[E_1^2]e^{iD(\Omega)z}
\end{align}
The total solution then becomes
\begin{align}\label{eq:SH-total}
B_2^{(t)}(z,\Omega)=ie^{iD_2(\Omega)z/2}\tfrac{\omega_1 d_{\rm eff}}{cn_2} z\FT[E_1^2]{\rm sinc}[D_2(\Omega)z/2]
\end{align}
If we now consider a transform-limited FW, $|\FT[E_1^2]|=\FT[|E_1|^2]$, then the SH intensity is $I_2(z,\Omega)= \tfrac{2\omega_1^2 d_{\rm eff}^2}{n_1n_2^2c^3\varepsilon_0} z^2{\rm sinc}^2[D_2(\Omega)z/2]I_1^2(\Omega)$. The classical result $I_2\propto {\rm sinc}^2(\Delta k z/2)$ is found by neglecting SH dispersion.

The "nonlocal" result from \cite{bache:2007a,bache:2008} that we used in Fig. \ref{fig:SH-Dk} can be recovered from $B_2^{(t)}$, as the nonlocal solution is precisely identical to the particular (or "driven wave" \cite{Valiulis:2011}) solution found above. This is because the nonlocal approach neglects the homogeneous solution ("free wave" \cite{Valiulis:2011}) but otherwise apply the same assumptions. In \cite{bache:2007a} a so-called nonlocal response function $R$ was introduced, and this approach corresponds to writing the particular solution (\ref{eq:SH-particular}) as
\begin{align}\label{eq:SH-particular-nonlocal}
B_2^{(p)}(\Omega)&=-\tfrac{\sqrt{2\pi}\omega_1 d_{\rm eff}}{cn_2 \Delta k} R(\Omega)\FT[E_1^2]\\
\label{eq:R-nonlocal}
R(\Omega)&=(2\pi)^{-1/2}\frac{\Delta k}{D_2(\Omega)}
\end{align}
The constants in the nonlocal response function $R$ are suitably chosen to normalize it properly. By analyzing $R$, \cite{bache:2007a,bache:2008} showed that when $d_{12}^2-2k_2^{(2)}\Delta k>0$ the response becomes resonant because the denominator has two real roots. For positive SH GVD (normal dispersion) this inequality is expressed as $\Delta k<\Delta k_{\rm r}^{\rm nl}$, where the threshold $\Delta k_{\rm r}^{\rm nl}\equiv d_{\rm 12}^{\rm 2}/2k_2^{\rm (2)}$ depends critically on GVM. It marks the transition between the non-resonant ($\Delta k>\Delta k_{\rm r}^{\rm nl}$) and resonant regime ($\Delta k<\Delta k_{\rm r}^{\rm nl}$). In the latter the resonance frequencies are
\begin{align}\label{eq:Wp-nl}
    \Omega^{\rm nl}_\pm=\left(d_{12}\pm [d_{12}^2-2\Delta k k_2^{(2)}]^{1/2}\right)/k_2^{(2)},
\end{align}
accurate to 2. order. Instead when $\Delta k>\Delta k_{\rm r}^{\rm nl}$ the nonlocal response is non-resonant and ultrabroadband; when including up to 2. order dispersion it assumes a Lorentzian shape with peak position at $\Omega^{\rm nl}=d_{12}/k_2^{(2)}$.

The traditional "phase-matched sidebands" theory uses the classical result $I_2\propto {\rm sinc}^2(\Delta k z/2)$, and phenomenologically generalizes to full chromatic dispersion $\Delta k(\omega)=k_2(\omega)-2k_1(\omega/2)$. In absence of phase matching, phase-matching can occur between a sideband frequency in the FW spectrum $\omega_1'$ and its corresponding SH sideband frequency $\omega_2'=2\omega_1'$. By expanding $\Delta k(\omega)$ around $\omega_2$, we see that when $\Delta k<\Delta k_{\rm r}^{\rm sb}\equiv d_{12}^2/[2k_2^{(2)}-k_1^{(2)}]$, phase-matching occurs at the SH frequency offsets
\begin{align}\label{eq:Wp-pm}
    \Omega_\pm^{\rm sb}&=\frac{d_{12}\pm
    [d_{12}^2-\Delta k(2k_2^{(2)}-k_1^{(2)})]^{1/2}
    }{k_2^{(2)}-k_1^{(2)}/2}
\end{align}

\begin{figure}[tb]
\centerline{\includegraphics[width=7cm]{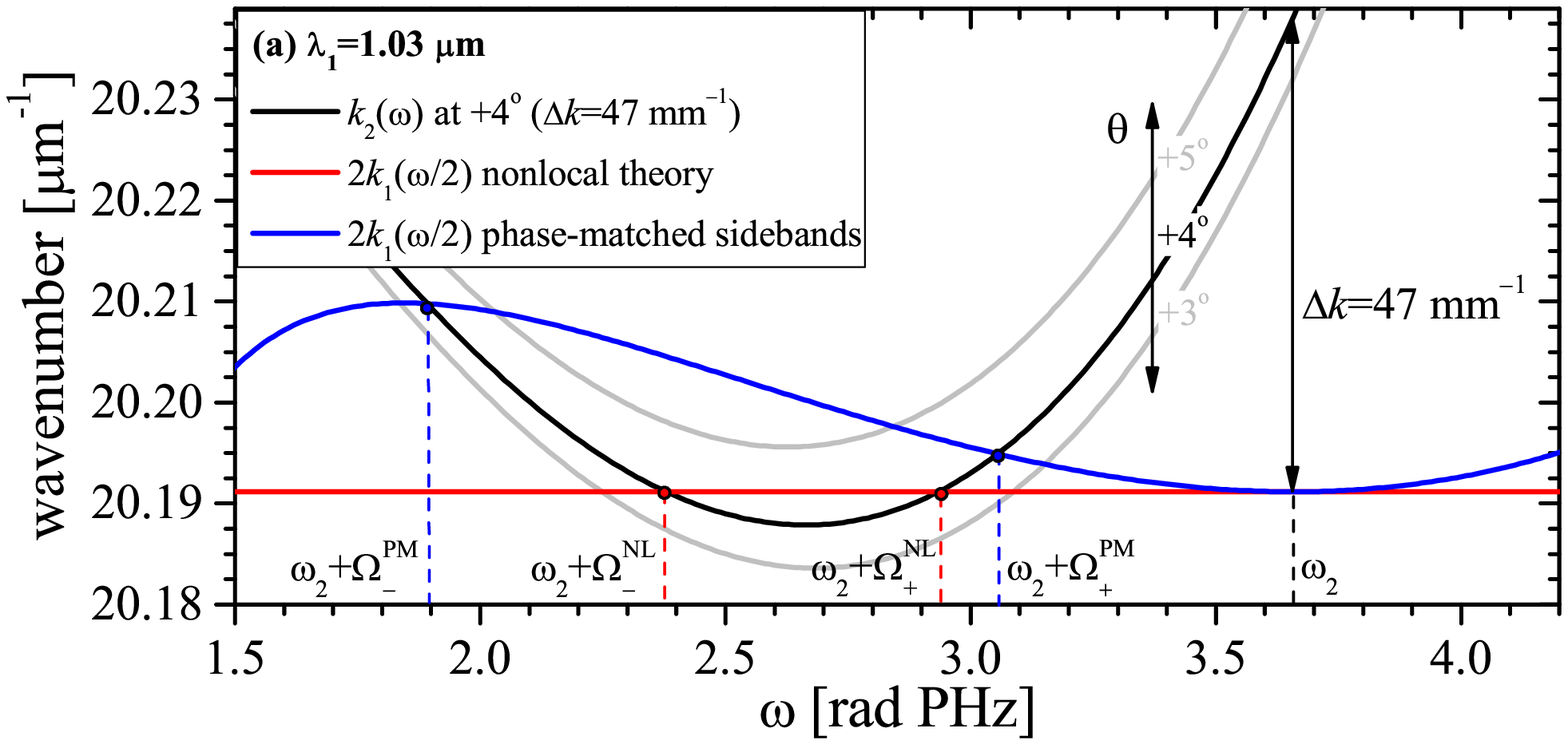}}
\centerline{\includegraphics[height=3cm]{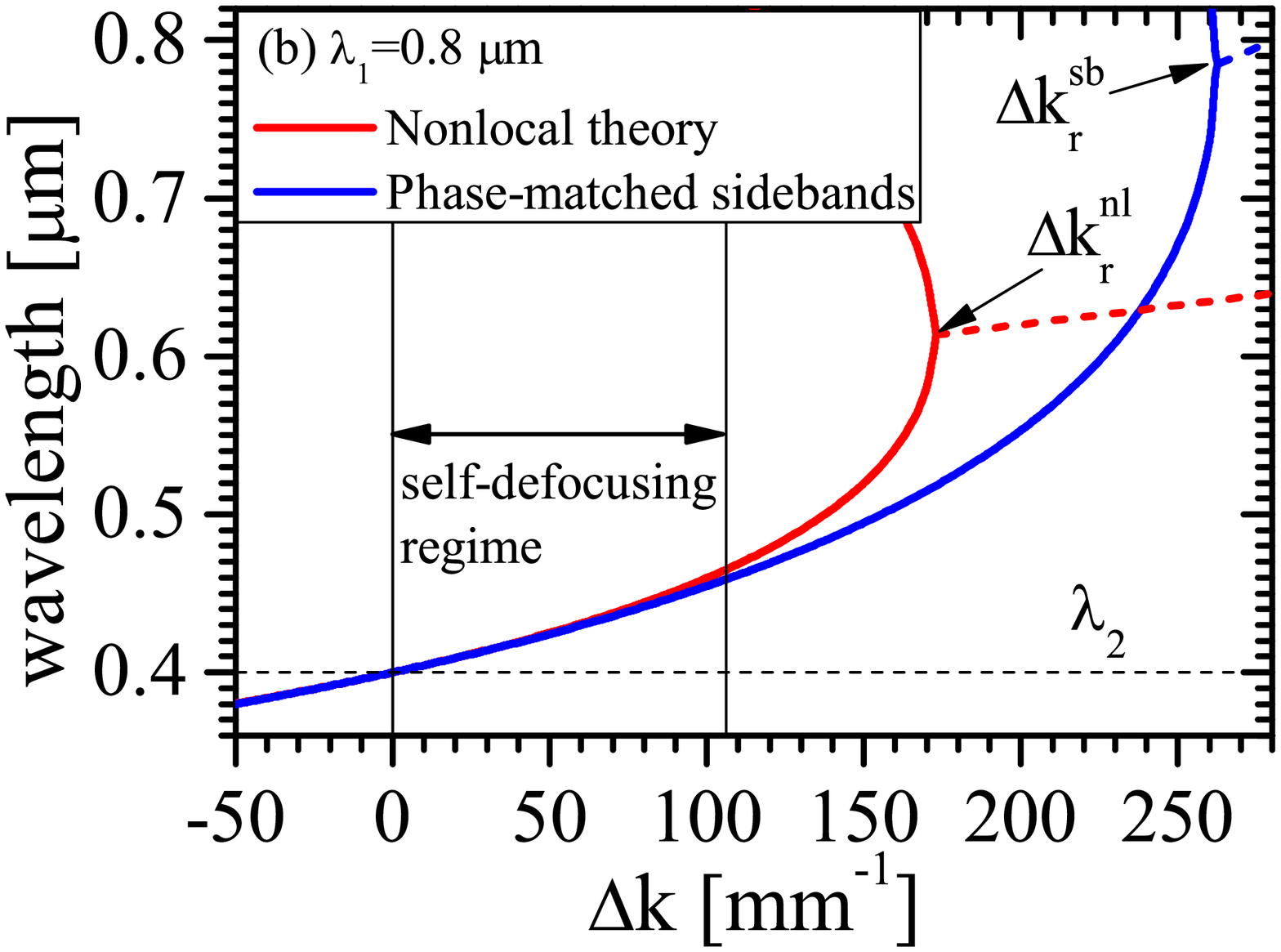}
\includegraphics[height=3cm]{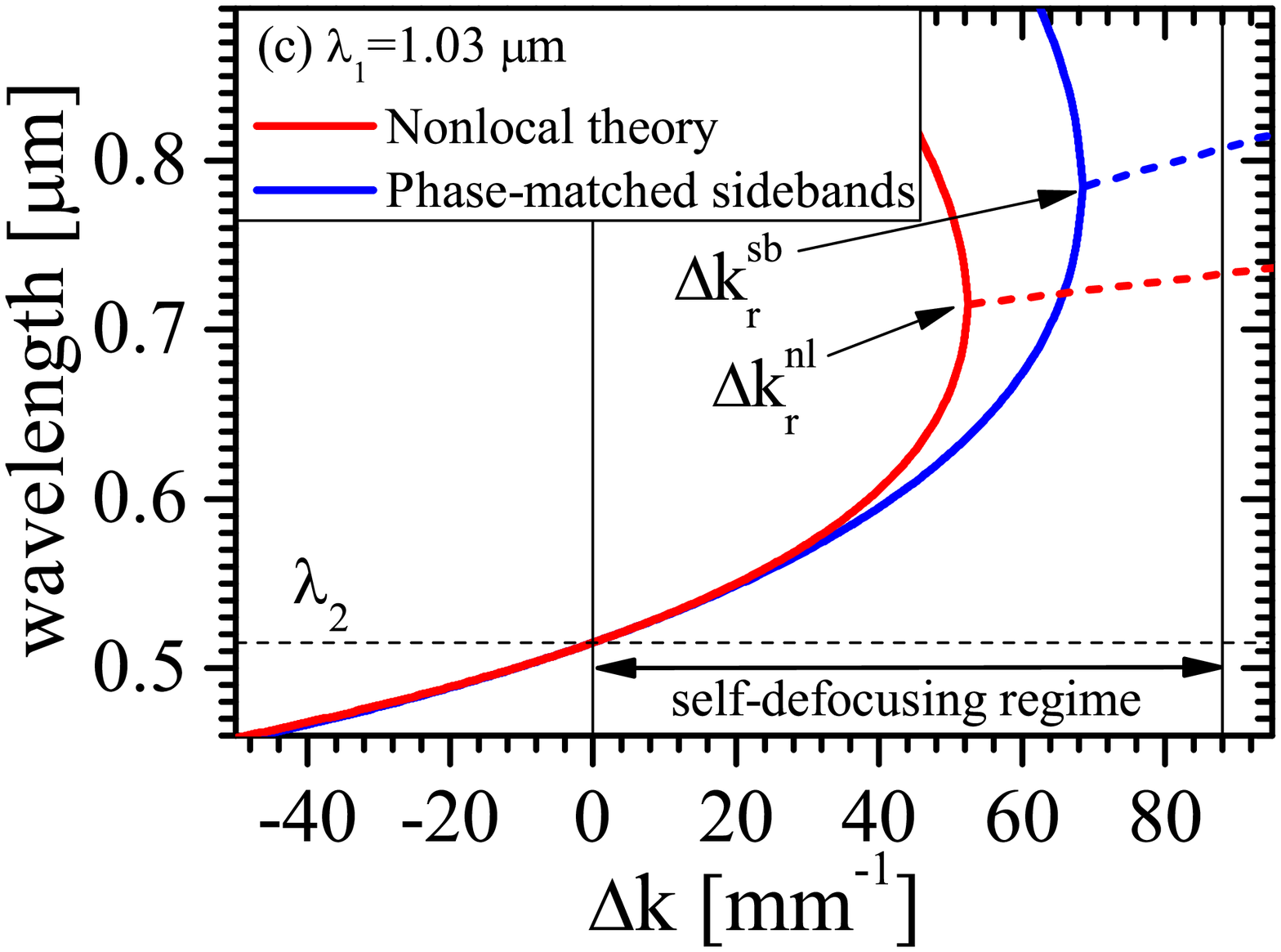}}
\caption{(Color online) The SH resonance wavelengths calculated for type I birefringent ($oo\rightarrow e$) SHG in BBO. 
(a) Illustration of the phase-matching condition $k_2(\omega)-2k_1(\omega/2)$ at $+4^\circ$ of Fig. \ref{fig:SH-Dk}, reported in the FW reference frame $\tau=t-zk_1^{(1)}$. The SH curve (black) is tunable through $\theta$; this is indicated with gray curves for $\theta$ taken $1^\circ$ larger and smaller. (b+c) Calculated SH peak wavelengths vs. phase mismatch $\Delta k$ for $\lambda_1=0.8\mic$ $1.03\mic$, respectively. Full lines are resonant wavelengths, while dashed lines are the peak values of the Lorentzian shape of the nonresonant response. The self-defocusing regime is indicated where $n_{2,\rm casc}^I+n_{2,\rm Kerr}^I<0$. }
\label{fig:theory}
\end{figure}

Eqs. (\ref{eq:Wp-nl}) and (\ref{eq:Wp-pm}) seem quite similar but the latter includes dispersion of the FW (here up to 2. order through the FW GVD). In the nonlocal case the FW is assumed dispersion-free; it is a consequence of requiring $\FT[E_1^2]$ to be independent in $z$. Fig. \ref{fig:theory}(a) shows a graphical representation of the two cases in a classical $\omega-k$ dispersion diagram representing the $+4^\circ$ case of Fig. \ref{fig:SH-Dk}: For the nonlocal theory the FW dispersion curve is obviously flat as it is dispersion-free (red curve), while the SH follows the material dispersion curve (black). Their intersection points give the nonlocal resonance (phase-matching) frequencies. The phase-matched sidebands theory assumes that the FW is dispersive (i.e. follows the material dispersion, shown with a blue curve), evidently giving different resonance frequencies when intersecting with the SH curve. The figure also shows graphically how the phase-mismatch is found as the distance between the FW and SH curves at $\omega_2$, as well as how angle-tuning shifts the SH curve up or down, while leaving the FW curves unchanged. The $+5^\circ$ curve also shows the example where the angle is tuned to a point where the nonlocal theory predicts no phase-matching (the SH and FW curves do not touch), and instead of a resonant nonlocal response one will here have a non-resonant ultrabroadband response. Figures \ref{fig:theory}(b) and (c) show the predicted SH phase-matching wavelengths vs. $\Delta k$ for $\lambda_1=0.8$ and $1.03\mic$. The two theories agree around $\Delta k=0$, but start to differ when $\Delta k \gg 0$. For $\lambda_1=0.8\mic$, (b) shows very strong separation of the two cases, but it only occurs in self-focusing regime [i.e. where the total nonlinear index change $\Delta n=(n_{2,\rm casc}^I+n_{2,\rm Kerr}^I)I_1>0$]. As BBO has normal GVD below $\lambda=1.4\mic$, solitons require a self-defocusing nonlinearity to exist. Instead for $\lambda_1=1.03\mic$, case (c), the deviation between the curves occur well in the self-defocusing regime that supports soliton formation. For higher $\lambda_1$ the two theories become more and more indistinguishable as the FW dispersion is reduced.

The experiment therefore used a commercial optical parametric amplifier to generate an $o$-polarized pump at $1.03\mic$ with ${40~\mu}$J pulse energy, with near-transform limited pulses (58 fs FWHM duration, inferred from an intensity autocorrelator, and a Gaussian-shaped 28.4 nm FWHM spectrum). The pump beam was collimated with an all-reflective telescope setup (beam spot size 0.5 mm FWHM). The input beam intensity was controlled by a neutral density filter. We used a 25-mm-long BBO crystal with a $10\times7~{\rm mm^2}$ aperture (cut with $\theta=21^{\circ}$, $ \phi=-90^\circ $). The phase-mismatch was tuned by rotating the external crystal angle with $1/6^{\circ}$ precision, and the total operational range of the external angle was kept low enough to avoid geometrical effects due to non-perpendicular incidence of the pump. We kept the input intensity fixed for all tuning angles; it must be intense enough to achieve a good signal of the detuned peak in the entire $\Delta k $ range. The cascading strength is $n_{2,\rm casc}^I\propto -d_{\rm eff}^2/\Delta k$, and while $d_{\rm eff}$ does not change much in the applied tuning range, the $1/\Delta k$ scaling gives severe changes in the cascading nonlinearity. We tried two different levels, $80~{\rm GW/cm}^2$ and $160~{\rm GW/cm}^2$, both with similar results. As peak splitting occasionally occurred with $160~{\rm GW/cm}^2$, in what follows we therefore present the $80~{\rm GW/cm}^2$ results. The SH spectrum was recorded by impinging the SH beam center on the spectrometer input fiber connector. We also used an integrating sphere, which gives a spatially averaged signal; this gave more blurred peaks indicating some spatial variation of the spectral contents, but the overall analysis and results presented in what follows are representative of both cases. Fig. 2(c) in \cite{Bache:2013} shows representative FW spectra recorded under similar conditions.

\begin{figure}[tb]
\centerline{\includegraphics[width=8.5cm]{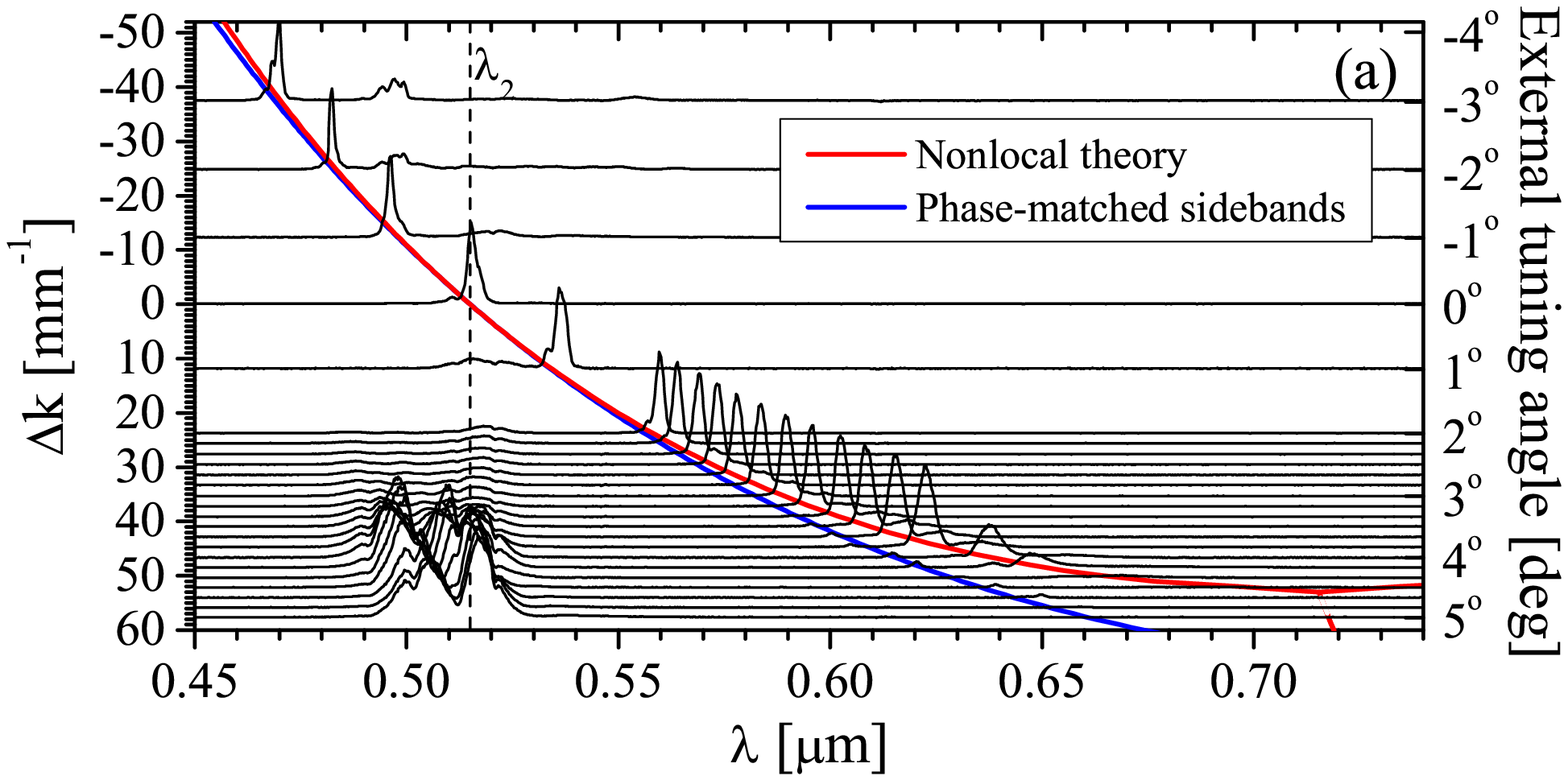}}
\centerline{\includegraphics[width=8.cm]{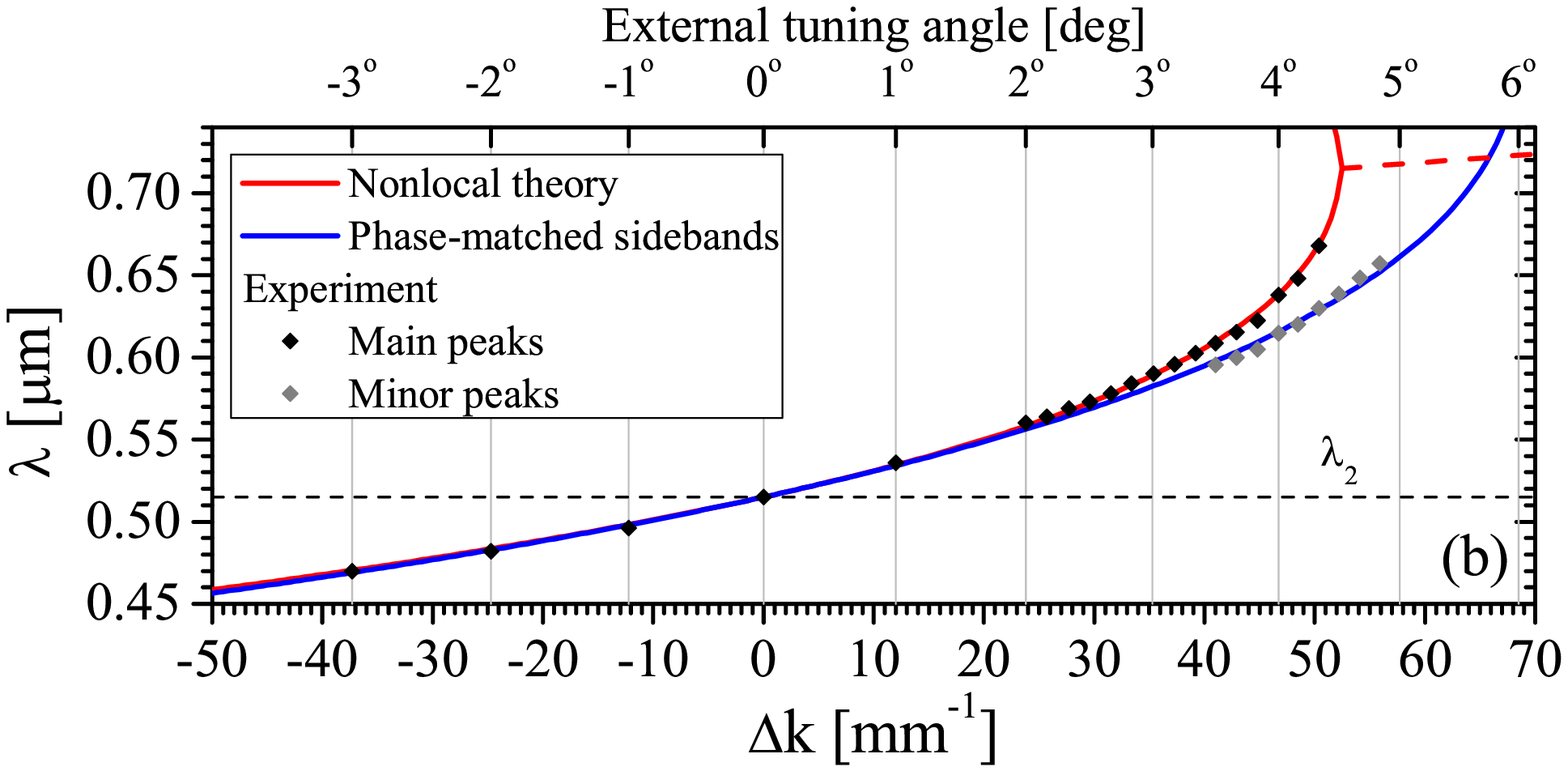}}
\caption{(Color online) (a) Experimental evolution of the SH spectrum vs. $\Delta k$; the results in Fig. \ref{fig:SH-Dk}(a) are selected from this data. All spectra are normalized to the same peak value. The baseline of each spectrum indicates the invoked phase mismatch. (b) Quantitative comparison between the theories and the experimentally measured SH peak wavelengths.
}
\label{fig:lambda-dk}
\end{figure}

Figure \ref{fig:lambda-dk}(a) shows the results of Fig. \ref{fig:SH-Dk}(a) in more detail. Around $\Delta k=0$ the SH detuned peak is dominating the spectrum but the two theories are almost identical there. At higher $\Delta k>30\imm$ it becomes clear that the dominating SH peak is best explained by the nonlocal theory (red curve). In this range some minor blue-shifted peaks are present, which seem to obey the phase-matched sidebands theory. As the transition to the non-resonant regime is approached ($\Delta k_{\rm r}^{\rm nl}\simeq 52\imm$) the peak becomes very broad and is now no longer dominating; around zero detuning ($\lambda_2$) some modulated peaks are instead dominating. We have extracted the wavelengths of the observed peaks and show in (b) a quantitative comparison between the two models. Clearly the nonlocal theory accurately predicts the major peaks observed in the SH spectrum, which indirectly proves that we have excited a FW self-defocusing soliton for $\Delta k>0$. Further evidence for this is that we in this regime observed moderate self-compression effects in autocorrelation traces of the FW, which is a typical feature of solitons . The quantitative comparison also reveals that the minor blue-shifted peaks indeed follow the phase-matched sidebands curve; evidently part of the FW is dispersive and excites these minor peaks. Finally, beyond the transition $\Delta k_{\rm r}^{\rm nl}\simeq 52\imm$ the detuned peak disappears because the nonlocal response becomes nonresonant. 

We could not observe the upper branch of the resonant response. It is quite elusive here as its intensity is $\propto |R(\Omega)|^2|\FT[E_1^2]|^2$, so a significant amount of spectral broadening of the FW is needed to see a signal far away from zero detuning $\lambda_2=0.515\mic$. This also explains why the SH nonlocal peak quickly decreases close to the transition $\Delta k_{\rm r}\simeq 52\imm$. On the other hand, for zero GVM the transition $\Delta k_{\rm r}^{\rm nl}=0$, and both branches will develop symmetrically around $\lambda_2$ for $\Delta k<0$ \cite{Zhu:2004}.


\begin{figure}[tb]
\centerline{\includegraphics[width=4.cm]{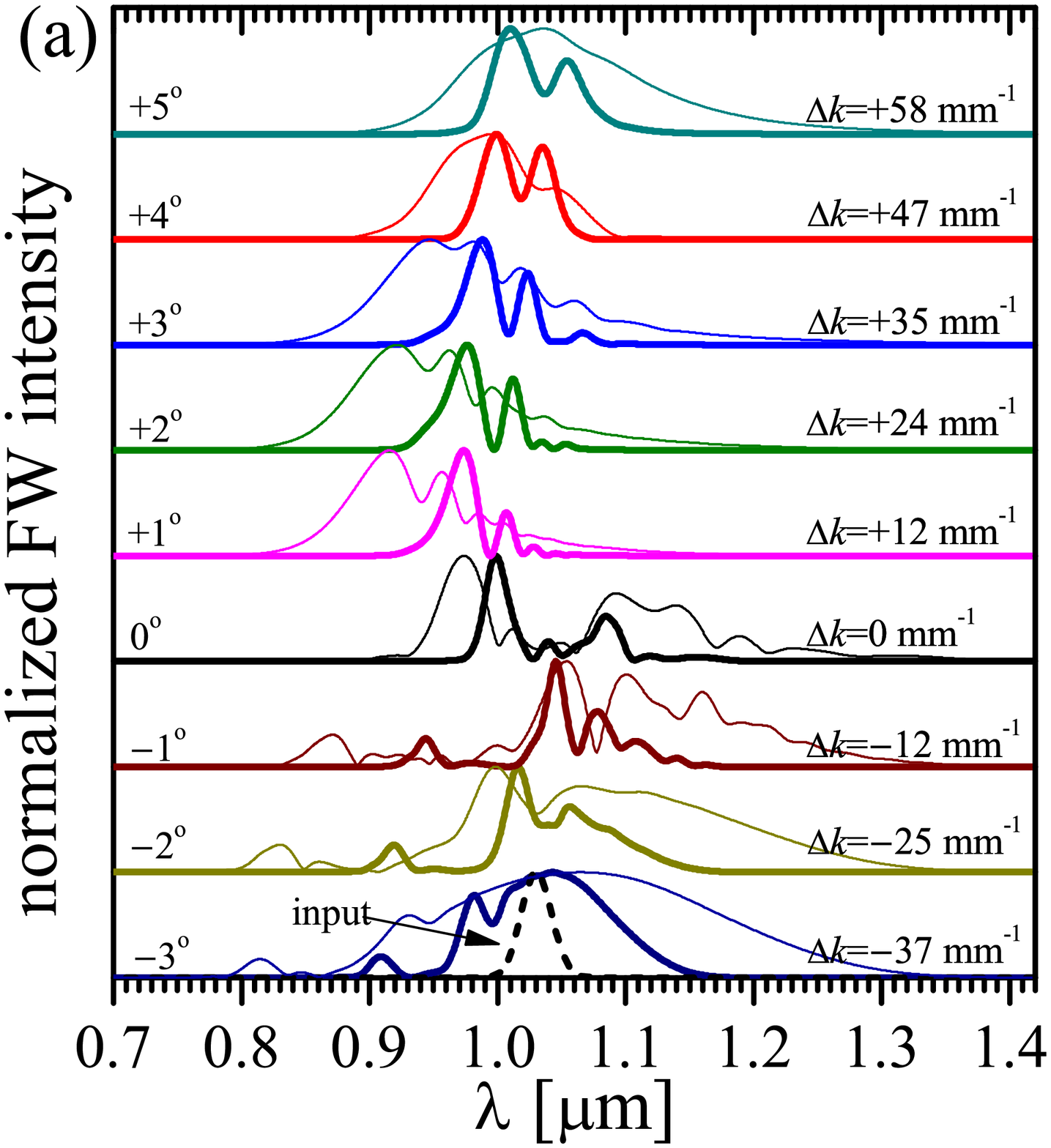}
\includegraphics[width=4.cm]{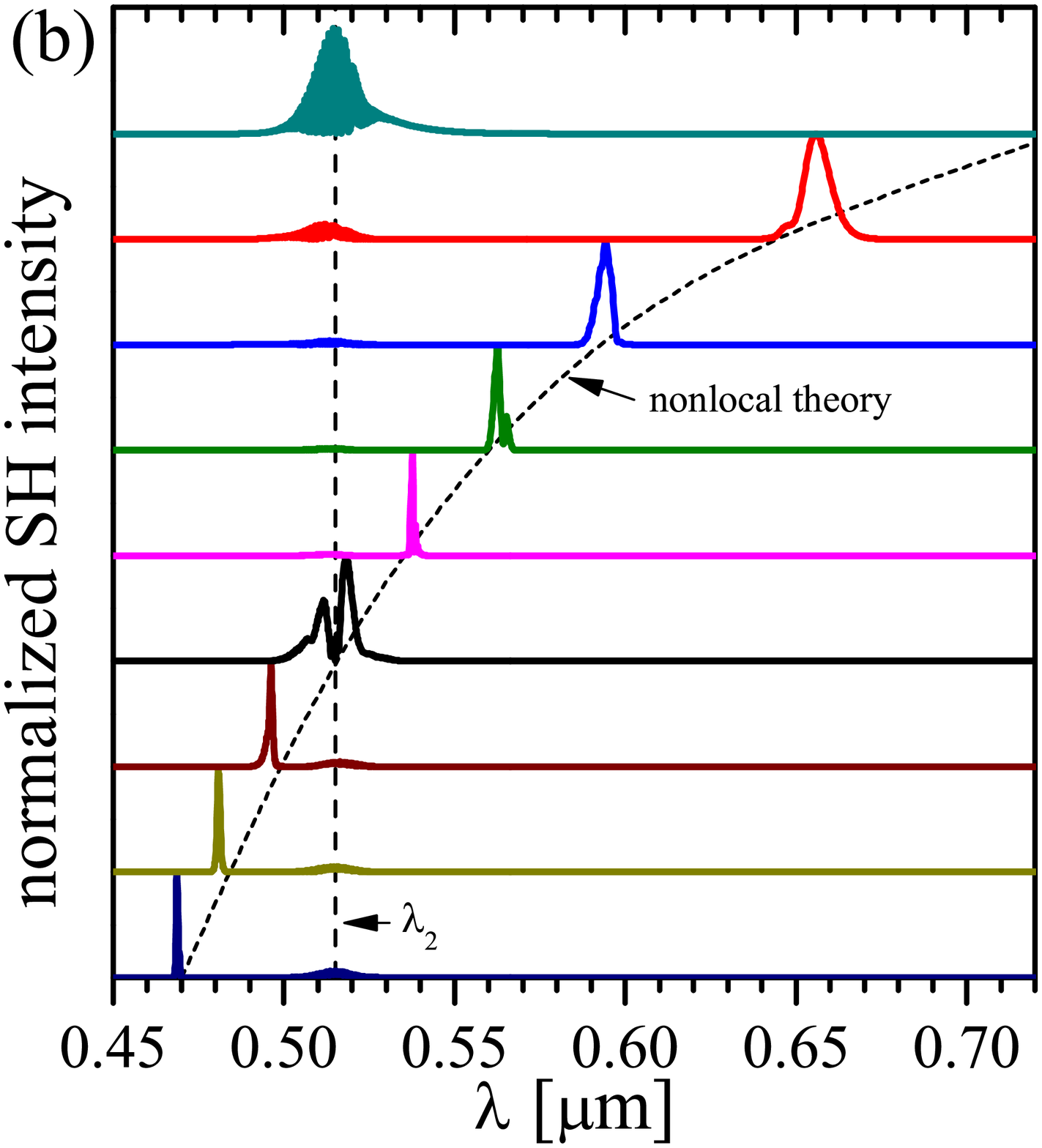}}
\centerline{\includegraphics[width=4.cm]{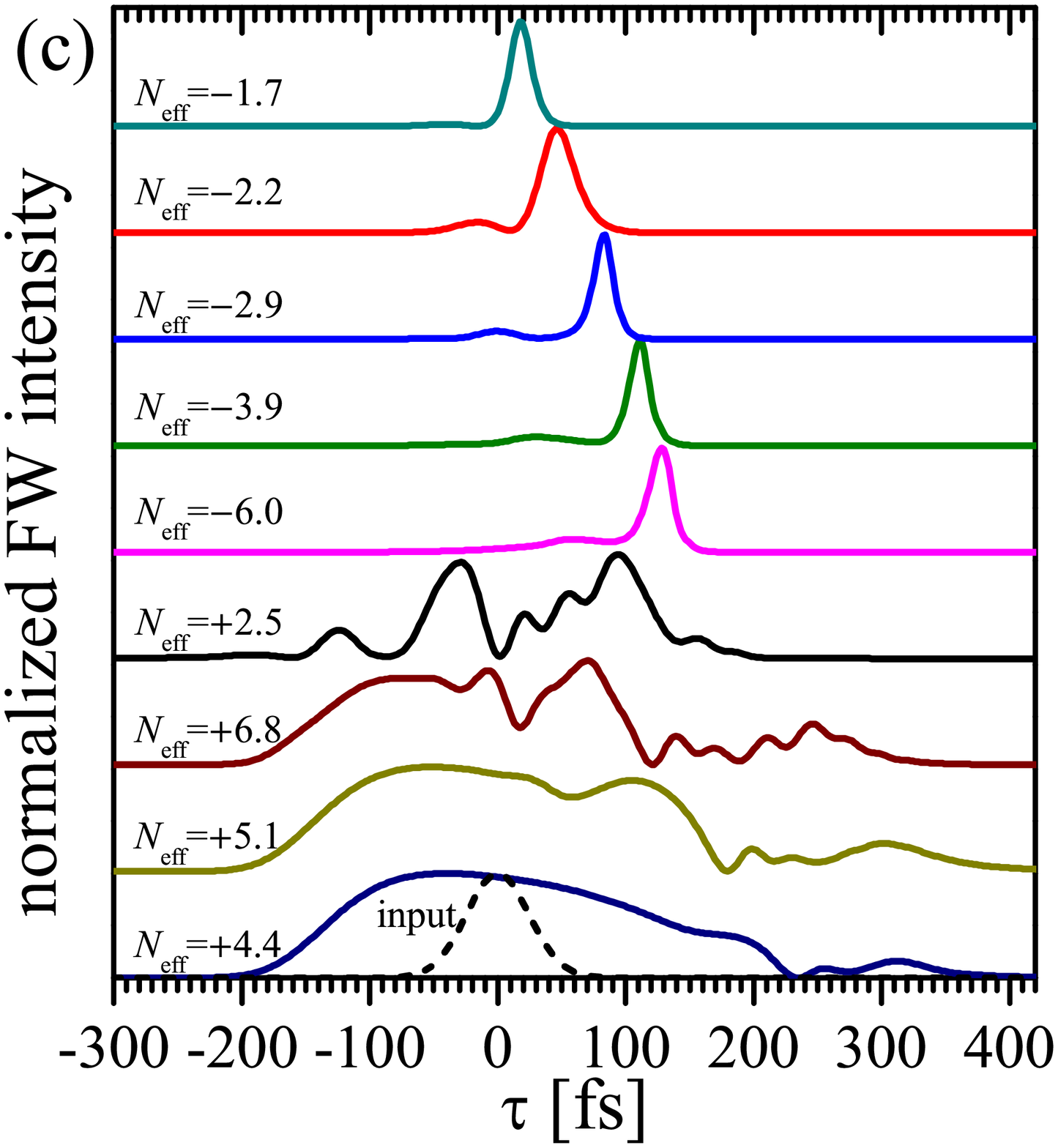}
\includegraphics[width=4.cm]{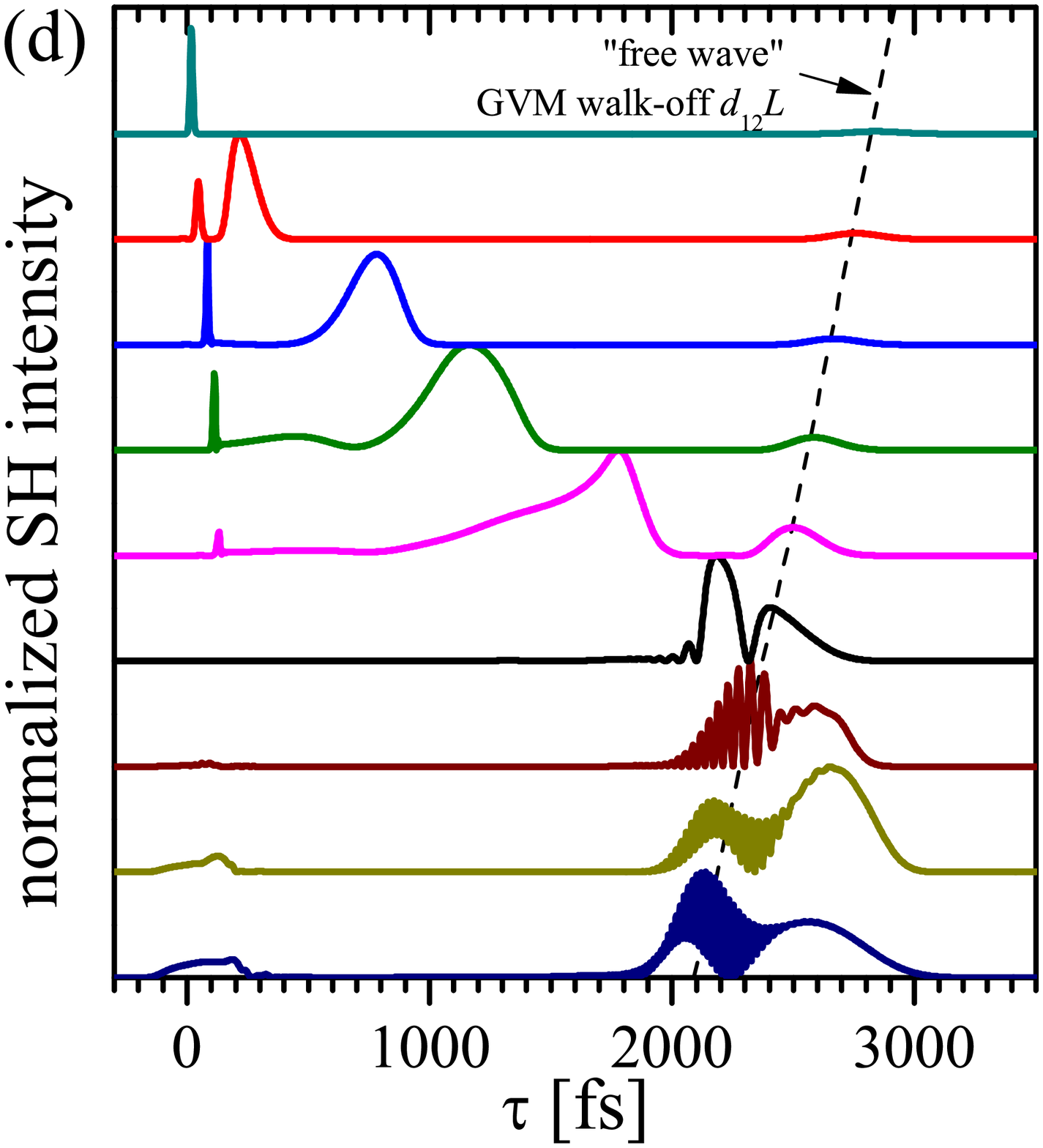}}
\centerline{\includegraphics[width=8.5cm]{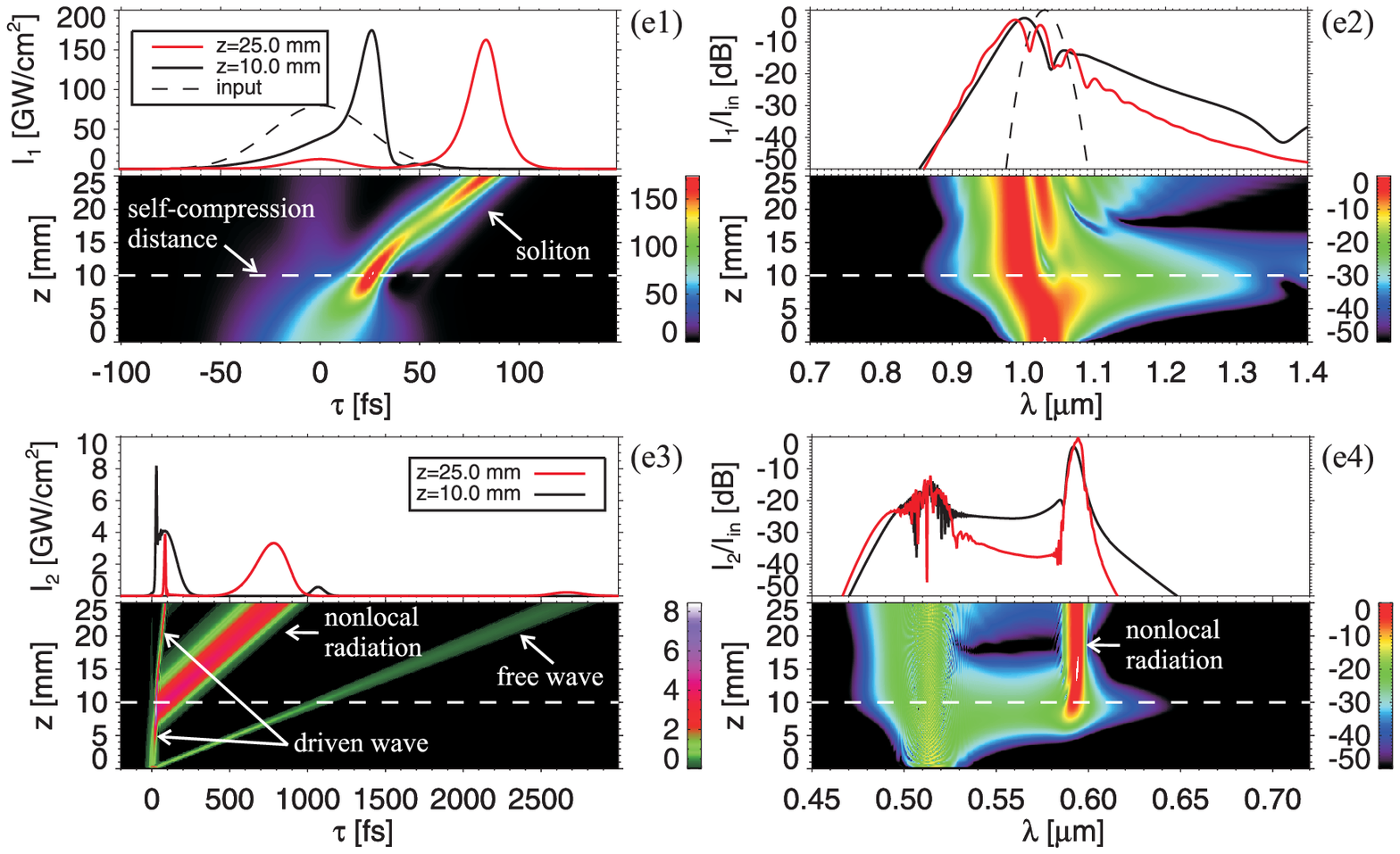}}
\caption{(Color online) Numerical slowly-evolving wave approximation \cite{moses:2006,bache:2007} simulations using experimental parameters. (a+b) Normalized spectra; in (a) $|\FT[E_1^2(\tau)]|$ is also shown (thin lines). (c+d) Normalized temporal intensities (notice the different time axis scalings). An offset is imposed between each curve for clarity of presentation.
(e) Propagation dynamics of $+3^\circ$-case.
Kerr nonlinearities are from \cite{Bache:2013}.}
\label{fig:sim}
\end{figure}

In Fig. \ref{fig:sim} we show the results of plane-wave numerical simulations with phase-mismatch values identical to Fig. \ref{fig:SH-Dk}. The FW spectra (thick lines) in (a) show significant spectral broadening. The SH spectral trend in (b) is very similar to the experiment [Fig. \ref{fig:SH-Dk}(a)]; in particular we verify that at $+5^\circ$ we see the same type of spectrum as in the experiment, namely that the radiation around $\lambda_2$ dominates due to the broadband nonlocal response. 
Since the theory predicts that $I_2(\Omega)\propto |R(\Omega)|^2|\FT[E_1^2(\tau)]|^2$, and not $I_2(\Omega)\propto |R(\Omega)|^2I_1^2(\Omega)$, we also show $|\FT[E_1^2(\tau)]|$ in (a) with thin lines: due to a significant FW phase it becomes much broader than $I_1(\Omega)$, and thus $I_1(\Omega)$ is not a good measure to predict $I_2(\Omega)$. Figures \ref{fig:sim}(c) and (d) show the time traces: For negative $\Delta k$ the total nonlinearity is self-focusing, and since the GVD is positive the FW broadens temporally (wave-breaking \cite{Tomlinson:1985}). At positive $\Delta k$ the total nonlinearity is self-defocusing and (c) shows that the intensity is large enough to excite a FW soliton, leading to a moderate compression (just like what we observed in the experiment). To quantify this the effective soliton order \cite{bache:2007} is shown in each case, calculated from the effective nonlinear Kerr index $n_{2,\rm eff}^I=n_{2, \rm casc}^I+n_{2,\rm Kerr}^I$, and the sign of the soliton order is intended to distinguish the self-focusing case ($n^I_{2,\rm eff}>0$) from the self-defocusing case ($n^I_{2,\rm eff}<0$). As the self-defocusing soliton orders employed are all larger than unity soliton self-compression is observed for all $\Delta k>0$ cases. A strong pulse shock front is seen for low positive $\Delta k$ that gradually degrades along with the pulse compression factor for higher $\Delta k$. This is because the cascading self-steepening term is $\propto d_{12}/\Delta k$ \cite{ilday:2004,moses:2006b,bache:2007a} and because the effective soliton order decreases, respectively. In (d) the SH temporal intensities are shown. The dashed line indicates the calculated walk-off delay due to GVM; this is the "free" wave. The "driven" wave appears in the vicinity of the $\tau=0$ regime; it is essentially a temporal copy of the FW (particularly evident for large $\Delta k$). The strong SH components observed in between the driven and free waves for positive $\Delta k$ are caused by the nonlocal resonance effect. To appreciate this, (e) shows the propagation dynamics of the $+3^\circ$ case as a descriptive example: the nonlocal radiation, observable both in the SH time trace in (e3) and spectral trace in (e4), only emerges \textit{after} the soliton self-compression point, i.e. only after the FW soliton is actually formed, which occurs at $z=10$ mm, see (e1). This is another evidence that it is truly a soliton-induced nonlocal resonance. Before the soliton forms, the FW is dispersive and thus follows the phase-matched sideband phase-matching condition; however no significant radiation is observable as the FW spectral broadening is too weak in this regime to excite the SH resonance with a detuning this large.

Instead of angle-tuning the crystal, a quasi-phase-matching (QPM) geometry could be used in ferroelectric crystals like lithium niobate, and the tunability of the SH peak is ensured with continuously variable grating period $\Lambda$ (fan-out)  (as done in, e.g. \cite{Moutzouris2006}). Our results remain unchanged: one can still find a resonant regime close to phase matching, and essentially in the nonlocal response function $\Delta k$ must be replaced by the effective QPM phase mismatch $\Delta k_{\rm QPM}=\Delta k-2\pi/\Lambda$, see \cite{zeng:2012}. However, when using QPM to reduce the effective phase-mismatch to a non-zero level (to achieve cascading) there are usually some higher-order QPM processes that become very nearly phase-matched. Therefore it is typical to observe numerous resonant peaks in the SH spectrum, see e.g. \cite{Guo:2013}.

In summary, phase-mismatched SHG of femtosecond pulses in a thick crystal leads to the formation of a dominating compressed peak in the SH spectrum. While this is well known, in contrast to earlier experiments we used a large input intensity. The peak was tunable over 100's of nanometers until it abruptly disappeared for large positive tuning angles (positive $\Delta k$). The traditional theory of phase-matching between a sideband in the broadband FW and a sideband of the SH fell short in explaining the results. Instead we considered a "nonlocal" theory, where the narrow wavelength-tunable SH peak is explained by a resonance in the nonlocal response in the cascading limit of strongly phase-mismatched SHG. In the nonlocal approach a key assumption is that the FW is dispersion-free, in contrast to the traditional explanation that phenomenologically considers a dispersive FW. The experimental conditions were carefully selected to discern the two theories: for positive tuning angles ($\Delta k>0$) they started to deviate, and we found that the nonlocal theory gave an excellent agreement to the observed peaks. This result provided significant evidence that we had excited a FW self-defocusing soliton: for $\Delta k>0$ a dominating self-defocusing nonlinearity is induced by the cascaded SHG, and since the pump wavelength was in the normal dispersion regime and a high intensity was used we could excite a self-defocusing soliton. A soliton is precisely characterized by being dispersion-free and thus accurately described by the nonlocal theory. The tunable and compressed SH peaks observed for positive tuning angles are therefore the soliton-induced nonlocal resonances that were predicted theoretically \cite{bache:2007a}. This is a completely new way of confirming the presence of a soliton in this elusive resonant nonlocal regime. The historical experiments \cite{bakker:1992,Rassoul:1997,Pioger:2002,Zhu:2004, baronio:2004,Moutzouris2006,Marangoni:2007,Pontecorvo2011} instead employed low intensities so solitons could not be excited. Thus the FW remained dispersive and therefore the traditional theory of phase-matched sidebands could accurately explain the peaks.

MB and BZ acknowledge support from the Danish Council for Independent Research, projects no. 274-08-0479 and no. 11-106702. Ole Bang and Xianglong Zeng are acknowledged for fruitful discussions.

\appendix

\section{The nonlocal theory and phase-matched sidebands theory}

Here we for completeness discuss separately the two theories as they traditionally have been presented, before we show a framework that combines the two approaches. The final comparison and discussion lay the grounds for the theoretical section in the main parts of the paper.

We consider SHG $\omega_1+\omega_1\rightarrow \omega_2$ between the FW (frequency $\omega_1$) and the SH (frequency $\omega_2=2\omega_1$). The plane-wave coupled SHG equations under the slowly-varying envelope approximation of the electric field envelopes $E_j$ are (in mks units)
\begin{align}\label{eq:shg-fw-a}
   \left[i\tfrac{\partial}{\partial z}- \tfrac{1}{2}k_1^{(2)}\tfrac{\partial^2}{\partial \tau^2}\right]E_1+\tfrac{\omega_1
   d_{\rm eff}}{cn_1} E_1^*E_2 e^{i\Delta k z}&=0\\
\label{eq:shg-sh-a}
\left[i\tfrac{\partial}{\partial z}-id_{12}\tfrac{\partial}{\partial
    \tau}- \tfrac{1}{2}k_2^{(2)}\tfrac{\partial^2}{\partial \tau^2}\right]E_2+\tfrac{\omega_1 d_{\rm eff}}{cn_2} E_1^2
e^{-i\Delta k z}&=0
\end{align}
where $k_j(\omega)=n_j(\omega)\omega/c$ are the wave numbers, $n_j(\omega)$ the frequency dependent refractive indices of the FW ($j=1$) and SH ($j=2$), $n_j\equiv n_j(\omega_j)$, $\Delta k=k_2^{(0)}-2k_1^{(0)}$ the phase mismatch parameter, $d_{12}=k_1^{(1)}-k_2^{(1)}$ the GVM parameter,  $k_j^{(2)}$ the group-velocity dispersion (GVD) coefficients. Generally $k_{j}^{(m)}=\tfrac{d^mk_j}{d\omega^m}|_{\omega=\omega_j}$ are the dispersion coefficients at the reference frequencies $\omega_j$; we here only include up to 2. order dispersion for simplicity but will later generalize to higher-order dispersion. We could also include self-steepening effects, but the analytical results we now present are unaffected by this. Kerr nonlinearities are also neglected for simplicity; the results presented are intended to investigate the SH dispersion and nonlinear properties when thick-crystal femtosecond SHG is operated under phase-mismatched interaction. This implies that the crystal length $L$ is on the order of 10 mm or more, so the strongly phase-mismatched (cascading) limit $\Delta kL\gg 2\pi$ is always fulfilled. This means that the SH conversion is weak, even for high FW intensities, and thus that Kerr self-phase and cross-phase modulation of the SH is insignificant. They are instead both very relevant for the FW, but this is not what we investigate here. Finally, $d_{\rm eff}$ is the effective $\chi^{(2)}$ nonlinearity.

\subsection{Nonlocal theory}
The nonlocal theory builds on the approach used in \cite{bache:2007a}. We first assume a heavily phase-mismatched SHG process ($|\Delta k|L\gg 2\pi$), allowing for the ansatz
\begin{align}\label{eq:ansatz}
E_2^{\rm nl}(z,\tau)=A(\tau)e^{-i\Delta k z}
\end{align}
that separates the $z$ and $\tau$ dependence. Inserting the ansatz in Eq. (\ref{eq:shg-sh-a}) gives the ordinary differential equation $\Delta k A(\tau)-id_{12}A'(\tau)- \tfrac{1}{2}k_2^{(2)}A''(\tau)+\tfrac{\omega_1 d_{\rm eff}}{cn_2} E_1^2=0$, where primes denote time derivatives. Introducing the Fourier transform $E_2(z,\Omega)=(2\pi)^{-1/2}\int_{-\infty}^\infty d\Omega e^{+i\Omega \tau}E_2(z,\tau)$, in Fourier domain we get $A(\Omega)\left[\Delta k -d_{12}\Omega+ \tfrac{1}{2}k_2^{(2)}\Omega^2\right]+\tfrac{\omega_1 d_{\rm eff}}{cn_2} \FT[E_1^2]=0$, implying the solution
\begin{align}\label{eq:SH-locked-a}
  E_2^{\rm nl}(z,\Omega)=-e^{-i\Delta k z}\sqrt{2\pi}\frac{\omega_1 d_{\rm
      eff}}{cn_2\Delta k} R(\Omega) \FT[E_1^2]
\end{align}
where $\FT[\cdot]$ denotes the forward Fourier transform, and we used Eq. (\ref{eq:ansatz}). A normalized (how it is normalized is discussed later) nonlocal response function is here introduced as
\begin{align}\label{eq:Rnonlocal-a}
  R(\Omega)=\frac{1}{\sqrt{2\pi}}\frac{\Delta k}{\tfrac{1}{2}k_2^{(2)}\Omega^2-d_{12}\Omega+\Delta k}
\end{align}
which turns out to be inherently dimensionless. We see from this result that the SH becomes "slaved" or "locked" to the FW: The SH spectral density for a transform-limited FW is therefore $I_2(\Omega)\propto|R(\Omega)|^2I_1^2(\Omega)$. For a chirped FW the relation becomes more complicated ($|\FT[E_1]^2|\neq |E_1(\Omega)|^2$), and one can no longer simply consider SH spectrum as a product of the FW spectral intensity and the nonlocal response function.

The ansatz Eq. (\ref{eq:ansatz}) reflects the strong cascading limit ($|\Delta k|L\gg 2\pi$), where it is assumed that the phase-mismatch is so large that the coherence length $\pi/|\Delta k|$ is much smaller than any other characteristic length scales (see also discussion in \cite{bache:2007a}). However, it turns out to work quite very well even when one of the other length scales become similar in size. An important length scale in this comparison is the quadratic nonlinear length scale defined through the traditional $\Gamma$-parameter $\Gamma=\omega_1d_{\rm eff}E_{1,\rm in}/(c\sqrt{n_1n_2})$, where $E_{1,\rm in}$ is the peak electric input field. The ansatz holds when $\Delta k\Gamma\gg 1$ \cite{Bache:2013}. The ansatz looks for solutions that are stationary in $z$, and this only happens when the FW can be assumed undepleted, but also when the FW phase does not change with $z$; remember from Eq. (\ref{eq:shg-sh-a}), where the "source term" $\FT[E_1^2]$ could induce SH variations in $z$ either through its amplitude or phase.

The denominator of the nonlocal response can become resonant when $\Delta k<\Delta k_{\rm r}$, where $\Delta k_{\rm r}\equiv d_{\rm 12}^{\rm 2}/2k_2^{\rm (2)}$ is an important phase-mismatch value that depends critically on the GVM parameter $d_{12}$. It marks the threshold between the non-resonant ($\Delta k>\Delta k_{\rm r}$) and resonant regimes $\Delta k<\Delta k_{\rm r}$. In the latter the resonant nonlocal behaviour occurs because denominator will have two real roots, leading to resonant peaks in $R$. These resonance frequencies are to 2. order
\begin{align}\label{eq:Wp-nl-a}
    \Omega^{\rm nl}_\pm=\left(d_{12}\pm
    \sqrt{d_{12}^2-2\Delta k k_2^{(2)}}\right)/k_2^{(2)}
\end{align}
This result was also found in \cite{Valiulis:2011} using a different approach, but essentially taking the same key assumptions; this will be discussed later. Instead when $\Delta k>\Delta k_{\rm r}$ the nonlocal response is non-resonant: the resonance peaks disappears and the nonlocal response is ultrabroadband. This is the optimal situation for few-cycle pulse compression \cite{bache:2007a,bache:2008} (see also recent discussion in \cite{zhou:2012}).

Let us now show how the cascading leads to a nonlocal Kerr-like nonlinearity: Using the convolution
theorem $E_2(z,\tau)= -e^{-i\Delta k z}\frac{\omega_1 d_{\rm
    eff}}{cn_2\Delta k} \int_{-\infty}^{\infty} {\rm d}s
\bar R(s)E_1^2(z,\tau-s)$, where $\bar{R}(\tau)=\IFT[R]$ is the inverse
Fourier transform of the response function. Note that Eq. (\ref{eq:Rnonlocal-a}) is defined so $\int_{-\infty}^\infty d\tau \bar R(\tau)=1$; this is the basis of the normalization discussed above. Inserting  $E_2(z,\tau)$ into Eq.~(\ref{eq:shg-fw-a}), we get that the FW obeys the following equation
\begin{multline}
  \label{eq:fh-shg-nlse-nonlocal-a}
  \left[i\tfrac{\partial}{\partial z}- \tfrac{k_1^{(2)}}{2}\tfrac{\partial^2}{\partial \tau^2}\right]E_1
  \\+\tfrac{3\omega_1}{8n_1c}\chi_{\rm casc}^{(3)}E_1^*\int_{-\infty}^{\infty} {\rm d}s
  \bar R(s)E_1^2(z,\tau-s) =0.
\end{multline}
The leading nonlinearity is now cubic, with $\chi_{\rm casc}^{(3)}=-8\omega_1d_{\rm eff}^2/(3cn_2\Delta k)$ being the Kerr-like nonlinear coefficient induced by cascading; equivalently this can be expressed as a Kerr-like nonlinear refractive index $n_{2,\rm casc}^I=-2\omega_1 d_{\rm eff}^2/(c^2\varepsilon_0 n_1^2n_2\Delta k)$. This nonlocal Kerr-like nonlinearity is non-resonant in frequency domain when $\Delta k>\Delta k_{\rm r}$ and resonant in frequency domain when $\Delta k<\Delta k_{\rm r}$. In the latter case the temporal nonlocal response function is oscillatory in time domain $\bar R(\tau)\propto \sin(|\tau|/t_1)$ with some characteristic oscillation time $t_1=2|\Omega^{\rm nl}_+-\Omega^{\rm nl}_-|^{-1}$. Finally, in the local limit where the FW spectrum is very narrow we can make the approximation that $R(\Omega)$ is constant inside the spectrum, and thus $E_1^*(z,\tau)\int_{-\infty}^{\infty} {\rm d}s \bar R(s)E_1^2(z,\tau-s)\simeq E_1(z,\tau)|E_1(z,\tau)|^2$. This is the instantaneous Kerr-like nonlinearity induced by cascading. It will be competing with the intrinsic material cubic nonlinearity $\chi^{(3)}_{\rm mat}$, and effectively the FW will experience a total nonlinear refractive index change $\Delta n=(n_{\rm casc}^I+n_{\rm Kerr}^I)I_1$, where $I_1$ is the FW intensity and  the material Kerr nonlinear refractive index is $n_{\rm Kerr}^I=3\chi^{(3)}_{\rm mat}/4\varepsilon_0 n_1^2 c$. When including the next order in the local-limit expansion an additional cascading-induced self-steepening term results \cite{bache:2007a,bache:2008}, equivalent to the term found in \cite{ilday:2004,moses:2006b,moses:2007a} using a perturbative approach.

\subsection{Phase-matched sidebands theory}
Let us now discuss the traditional approach based on phase-matched sidebands: In absence of phase matching between the center frequency $\omega_1$ and $\omega_2$, phase-matched SHG of a pulsed beam can occur using a sideband frequency in the FW spectrum $\omega_1'$, which then generates a SH at $\omega_2'=2\omega_1'$ that is detuned from $\omega_2$. This explains the detuned SH peak observed. Obviously changing the amount of phase mismatch of the center frequencies changes the SH detuned frequency that can be phase-matched also changes, which explains the tunability.

In order to quantify this, it is well known that under the undepleted FW approximation ($I_1$ is constant in $z$) the SH intensity obeys the equation
\begin{align}\label{eq:SH-sinc}
    I_2(z)=\frac{2\omega_1^2 d_{\rm eff}^2}{n_1^2n_2c^3\varepsilon_0}z^2{\rm sinc}^2(\Delta kz/2)I_1^2
\end{align}
which can be derived directly from Eq. (\ref{eq:shg-sh-a}) by integrating over $z$ and neglecting SH dispersion; the "undepleted FW" assumption also implies that the FW does not depend on $z$, and therefore can be taken constant in this integration. We now study the chromatic variation of the phase-matching condition $\Delta k=k_2^{(0)}-2k_1^{(0)}$, i.e. $\Delta k(\omega)=k_2(2\omega)-2k_1(\omega)$: when $\Delta k(\omega_1)\neq 0$ we may look for a phase-matching point detuned away from the FW frequency $\omega_1'$, but where frequency conservation is obeyed $\omega_2'=2\omega_1'$. Graphically, the sinc-function at phase matching is centered at $\omega_2$, but in absence of phase matching it is shifted to a new frequency $\omega_2'$. Focusing on a sideband $\Omega^{\rm sb}$ to the SH frequency this can be written as $\Delta k(\Omega^{\rm sb})=k_2(\omega_2+\Omega^{\rm sb})-2k_1(\omega_1+\Omega^{\rm sb}/2)$, and expanding it we get $\Delta k(\Omega^{\rm sb})=\Delta k-d_{12}\Omega^{\rm sb}  +\frac{1}{2}{\Omega^{\rm sb}}^2(k_2^{(2)}-k_1^{(2)}/2)+O({\Omega^{\rm sb}}^3)$. When $d_{12}^2-2\Delta k(k_2^{(2)}-k_1^{(2)}/2)>0$, phase-matching occurs at the frequency offsets
\begin{align}\label{eq:Wp-pm-a}
    \Omega_\pm^{\rm sb}&=\frac{d_{12}\pm
    \sqrt{d_{12}^2-2\Delta k(k_2^{(2)}-k_1^{(2)}/2)}}{k_2^{(2)}-k_1^{(2)}/2}
\end{align}
accurate up to 2. order. We see that the condition for having phase-matched sidebands, $d_{12}^2-2\Delta k(k_2^{(2)}-k_1^{(2)}/2)>0$, is reminiscent of the resonance condition we employed for the nonlocal theory. If we express it through the phase-mismatch parameter it becomes $\Delta k<\Delta k_{\rm r}^{\rm sb}=d_{\rm 12}^{\rm 2}/(2k_2^{\rm (2)}-k_1^{(2)})$ when $k_2^{(2)}-k_1^{(2)}/2>0$. Note that an analytical result for the detuned SH frequency was calculated previously \cite{bakker:1992,Rassoul:1997} taking into account only GVM (i.e. accurate to 1. order only).

This approach is quite phenomenological, because the undepleted FW result Eq. (\ref{eq:SH-sinc}) is based on the monochromatic phase mismatch $\Delta k=k_2^{(0)}-2k_1^{(0)}$. Any direct influence of the FW dispersion is absent from this equation, and can therefore not play a role in the analysis. However, it intuitively makes sense that we may "track" the phase-mismatch variation vs. frequency and therefore make the generalization to $\Delta k(\omega)=k_2(2\omega)-2k_1(\omega)$. The nonlocal approach also highlights how easily the SH higher-order dispersion is taken into account to describe dispersion beyond the monochromatic phase-mismatch parameter, and in fact a sinc-like result can be derived in the case where SH dispersion is present (see \cite{Valiulis:2011}).

\subsection{The approach of Valiulas et al.}
To see this, we follow the approach in \cite{Valiulis:2011}. The first step is standard, and introduces an auxiliary SH amplitude $E_2(z,\tau)=B_2(z,\tau)e^{-i\Delta kz}$ so the SH equation becomes
\begin{align}\label{eq:SH-Valiulis}
   \left[i\tfrac{\partial}{\partial z}+\Delta k-id_{12}\tfrac{\partial}{\partial
    \tau}- \tfrac{1}{2}k_2^{(2)}\tfrac{\partial^2}{\partial \tau^2}\right]B_2+\tfrac{\omega_1 d_{\rm eff}}{cn_2} E_1^2=0
\end{align}
In Fourier domain this can be written as
\begin{align}\label{eq:SH-Valiulis-FD}
   \tfrac{\partial}{\partial z}B_2(z,\Omega)-i D_2(\Omega)B_2(z,\Omega)=i\tfrac{\omega_1 d_{\rm eff}}{cn_2} \FT[E_1^2]
\end{align}
where $D_2(\Omega)=\tfrac{1}{2}k_2^{(2)}\Omega^2-d_{12}\Omega+\Delta k$ is the effective SH dispersion operator in frequency domain. To solve this we first look for solutions to the homogeneous equation $\tfrac{\partial}{\partial z}B_2(z,\Omega)-i D_2(\Omega)B_2(z,\Omega)=0$, which are on the form $B_2^{(h)}=ae^{iD_2(\Omega)z}$. Under the assumption that $\FT[E_1^2]$ does not depend on $z$ a particular solution $B_2^{(p)}$ can be found that is constant in $z$. This makes the $\partial B_2^{(p)}/\partial z=0$, and therefore we directly get from Eq. (\ref{eq:SH-Valiulis-FD})
\begin{align}\label{eq:b2p}
    B_2^{(p)}(\Omega)=-\tfrac{\omega_1 d_{\rm eff}}{cn_2 D_2(\Omega)} \FT[E_1^2]
\end{align}
Inserting the total solution as a linear combination $B_2=B_2^{(h)}+B_2^{(p)}$ into Eq. (\ref{eq:SH-Valiulis-FD}) and using the boundary condition $B_2(z=0,\Omega)=0$ we get $a=-B_2^{(p)}(\Omega)$, and thus
\begin{align}\label{eq:b2h}
    B_2^{(h)}(z,\Omega)=\tfrac{\omega_1 d_{\rm eff}}{cn_2 D_2(\Omega)} \FT[E_1^2]e^{iD_2(\Omega)z}
\end{align}
The total solution then becomes
\begin{align}\label{eq:b2-total}
    B_2(z,\Omega)&=\tfrac{\omega_1 d_{\rm eff}}{cn_2 D_2(\Omega)} \FT[E_1^2]\left[e^{iD_2(\Omega)z}-1 \right]
    \\
    &=ie^{iD_2(\Omega)z/2}\tfrac{\omega_1 d_{\rm eff}}{cn_2} z\FT[E_1^2]{\rm sinc}[D_2(\Omega)z/2]
\end{align}
If we now consider a transform-limited FW, $|\FT[E_1^2]|=\FT[|E_1|^2]$, then we get
\begin{align}\label{eq:b2-total-int-sinc}
    I_2(z,\Omega)&=\tfrac{2\omega_1^2 d_{\rm eff}^2}{n_1n_2^2c^3\varepsilon_0} z^2{\rm sinc}^2[D_2(\Omega)z/2]I_1^2(\Omega)
\end{align}
where the usual sinc-like behavior is recovered, only here the full SH dispersion is present in the argument.

We also mention that the nonlocal result is recovered in this process: the particular solution (\ref{eq:b2p}) is namely exactly equivalent to the nonlocal result derived in Eq. (\ref{eq:SH-locked-a}). This is also what is denoted as the "driven" wave in \cite{Valiulis:2011}, and it should not be surprising that the particular solution found by assuming a $z$-independent behavior is identical to the nonlocal result. Instead the homogeneous part, denoted as the "free" wave in \cite{Valiulis:2011}, is neglected in the nonlocal approach, but it describes the well-known temporal walk-off wave that travels away from the FW after one GVM length $T_0/|d_{12}|$ and moves with the SH group velocity (and evidently also becomes affected by the SH HOD, cf. the $e^{iD_2(\Omega)z}$ phase term).

Still the FW dispersion is elusive in these approaches. We here note that the FW dispersion can indirectly affect the SH through the FW "source term", $\FT[E_1^2]$. Even when the undepleted FW assumption holds, a phase can namely be accumulated due to dispersion. This does make the source term $z$-dependent, and in order to show the consequence a more rigorous analysis is required. An example is found in Ref. \cite{Su:2006}, where they for simplicity keep the SH dispersion absent (thus no GVM or SH GVD). In the weak FW GVD regime (they use this approximation because they need to keep the FW amplitude constant in $z$ allowing only the phase to change in order to solve it analytically; in this approximation they are therefore neglecting that the GVD induces a decreasing peak FW intensity as the FW is spread out temporally) they derive a result similar to Eq. (\ref{eq:SH-sinc}) where in the sinc-term the $\Delta k$ term has a contribution from the FW GVD in exactly the same way as by expanding the phase mismatch to include higher-order dispersion terms. This indicates that the phenomenological expansion of the phase-mismatch parameter is correct.

We can be a bit more specific. Consider the case where the FW pulse, initially Gaussian $E_1(z=0,\tau)=E_{1,\rm in}e^{-t^2/2T_0^2}$, becomes affected by GVD. In the Fourier domain the initial pulse is $E_1(z=0,\Omega)=E_{1,\rm in}T_0 e^{-t^2T_0^2/2}$. The GVD manifests itself through the buildup of a quadratic phase, so the pulse becomes
\begin{align}\label{eq:E1-GVD-FT}
    E_1(z,\Omega)=E_{1,\rm in}T_0 \exp\left[-\Omega^2T_0^2/2+izk_1^{(2)}\Omega^2/2\right]
\end{align}
Inverse Fourier transforming we get
\begin{align}\label{eq:E1-GVD}
    E_1(z,\tau)=\frac{E_{1,\rm in}}{\sqrt{1-i zk_1^{(2)}/T_0^2}}\exp\left[\frac{-t^2}{2T_0^2(1-i zk_1^{(2)}/T_0^2)}\right]
\end{align}
and evaluating the intensity we get $I_1(z,\tau)=\frac{I_{1,\rm in}}{\sqrt{1+z^2/L_D^2}} e^{-t^2/(1+z^2/L_D^2)}$, where $L_D=T_0^2/|k_1^{(2)}|$ is the FW dispersion length; this is the classic result that as the pulse propagates along $z$ it spreads out in time and its peak intensity drops correspondingly. The peculiar feature is that in the Fourier domain the amplitude is not affected by $z$, only the phase is, but this property is unfortunately not conserved once we square the field and Fourier transform it. Instead we get
\begin{align}\label{eq:E12-GVD-FT}
    \FT[E_1^2]=\frac{E_{1,\rm in}^2T_0}{\sqrt{2(1-i zk_1^{(2)}/T_0^2)}} \exp\left[-\Omega^2T_0^2/4+izk_1^{(2)}\Omega^2/4\right]
\end{align}
It is exactly the amplitude prefactor $[2(1-i zk_1^{(2)}/T_0^2)]^{-\tfrac{1}{2}}$ that violates the assumption about a stationary ($z$-independent) source term. However, for a small FW GVD, i.e. when $z\ll L_D$, we can neglect the variation of the amplitude. Then we readily see that an additional phase $e^{+izk_1^{(2)}\Omega^2/4}$ is added to the source term. We therefore introduce a new auxiliary function $B_2=C_2e^{+izk_1^{(2)}\Omega^2/4}$, and the analysis from Eq. (\ref{eq:SH-Valiulis-FD}) remains the same, i.e. we solve for $C_2$ and replace $D_2(\Omega)\rightarrow D_2(\Omega)-k_1^{(2)}\Omega^2/4=\tfrac{1}{2}(k_2^{(2)}-k_1^{(2)}/2)\Omega^2-d_{12}\Omega+\Delta k$.

This shows that expansion of the $\Delta k$ part in the sinc-term of the classical SH conversion result corresponds to allowing the FW dispersion to change the phase of the FW field but not the amplitude, which must remain unchanged. We also note that the slaved/driven solution, i.e. what the nonlocal theory actually derives, will also be affected by such a FW GVD phase.

It seems clear from these considerations that the only difference between the two cases is the assumption posed on the FW. The phase-matched sidebands theory assumes that the FW follows the material dispersion. The nonlocal theory as a starting point assumes that the FW is not affected by higher-order dispersion, and only carries a phase and a group velocity. However, in both cases any effect of the FW dispersion to the theoretical result must come from releasing this assumption. In both cases this is handled by the same approximation, so at the end it is just a matter of choice whether the FW GVD will play a factor or not, and not a restriction. A well-known example of a FW that does not disperse nor change its amplitude in $z$, is the temporal soliton; it is merely described by a phase and a group velocity, and no other higher-order dispersion terms. The nonlocal theory is therefore a soliton-based approach while the traditional phase-matched sidebands theory is a dispersive (non-solitonic) approach.

We finally note that one can obviously generalize the frequency resonance derivations as well as the discussions above to include higher-order dispersion, but this will eventually require semi-analytical or numerical solutions. In the theoretical curves presented in the paper we use "exact" dispersion, where no polynomial expansion is used and the refractive indices used are from the material Sellmeier equations. The solutions are therefore found numerically.

\end{document}